\documentclass[%
 reprint,
 amsmath,amssymb,
 aps,
]{revtex4-2}

\usepackage{ytableau}
\usepackage[hyperfootnotes=false]{hyperref}
\hypersetup{
 colorlinks=true,
 citecolor=blue,
 linkcolor=red,
 urlcolor=blue}
\usepackage{mathtools}
\usepackage{nccmath} 
\usepackage{xcolor}
\usepackage{caption}
\usepackage{subcaption}
\usepackage{graphicx}
\usepackage{dcolumn,physics}
\usepackage{tabularx}
\usepackage{bm}

\begin{document}

\preprint{APS/123-QED}

\title{Fully Heavy Pentaquarks: Spectral Analysis, Production Dynamics, and Strong Two-Body Decays via the Extended Gursey–Radicati Formalism}
\author{Ankush Sharma, Alka Upadhyay}
 \affiliation{Department of Physics and Material Science, Thapar Institute of Engineering and Technology, Patiala, India, 147004}
 
\email{ankushsharma2540.as@gmail.com}

\date{\today}

\begin{abstract}
The study of exotic multi-quark states has garnered significant attention recently, particularly in heavy-quark dynamics within quantum chromodynamics (QCD). We perform a comprehensive spectroscopic analysis of fully heavy pentaquark states with quark configurations $cccc\bar{c}$ and $bbbb\bar{b}$, considering spin-parity quantum numbers $J^P = 1/2^-$, $3/2^-$, and $5/2^-$. We construct the color-spin wavefunctions to explore the internal structure and mixing effects in these exotic states. Using an extended form of the Gursey-Radicati mass formula by incorporating spin-dependent interactions, we calculated their mass spectra. The modification incorporates effective mass contributions and hyperfine interactions to improve the predictive power for these hadronic states. We systematically analyze their quantum numbers, including spin parity, isospin, and the eigenvalues of the quadratic Casimir operator, which characterize their symmetry properties. The calculated mass spectra are compared with existing theoretical predictions to assess the stability and possible decay channels of these states. The calculated mass spectra exhibit a strong dependence on the interplay between spin interactions and color configurations, shedding light on the binding mechanism within these fully heavy multiquark systems. To gain further insights into their stability and decay properties, we investigated their potential production modes from $b$-hadron decays. Our analysis identifies dominant strong decay channels, providing critical theoretical benchmarks for distinguishing these states in future LHCb or EIC experiments. This study offers new insights into the role of heavy-quark dynamics in exotic hadron spectroscopy, serving as a stringent test for effective QCD-based models and lattice QCD predictions.
\end{abstract}

\maketitle


\section{Introduction}
The discovery of exotic hadrons has profoundly expanded our understanding of Quantum chromodynamics (QCD) and the intricate dynamics of strong interactions. Among these exotic states, pentaquarks have garnered significant interest due to their unique quark compositions and the potential insights they provide into multi-quark interactions. Specifically, pentaquarks containing five heavy quarks (charm or bottom quarks) present a compelling case for study. Due to their relatively large masses compared to light quarks, heavy quarks simplify certain aspects of the strong interaction calculations. Studying fully heavy pentaquarks is crucial for understanding the strong interaction in the heavy-quark sector, where non-perturbative QCD effects dominate. These states provide a unique platform to test theoretical models of multiquark dynamics, including diquark correlations, heavy-quark symmetry, and color confinement mechanisms. Unlike conventional hadrons, fully heavy pentaquarks are expected to exhibit distinct binding properties and decay patterns, offering new insights into exotic hadron spectroscopy. Their investigation also complements ongoing experimental searches at facilities like LHCb, helping to refine our understanding of exotic states in QCD. In recent years, remarkable progress has been made in experimental and theoretical research of pentaquarks. The LHCb collaboration observation of pentaquark states in the $J/\psi p$ channel provided the first unambiguous evidence for the existence of these exotic states. In recent times, LHCb discovered hidden-charm pentaquark structures with a good level of statistical significance, which is way above the standard requirement of 5 $\sigma$. These states are $P_{\psi s}^\Lambda(4338)^0$, $P_{cs}(4459)$, $P_c(4312)^+$, $P_c(4450)^+$, etc \cite{20211278,4338,PhysRevLett.115.072001, PhysRevLett.117.082003,PhysRevLett.117.082002}. These discoveries sparked a wave of theoretical models attempting to describe the observed pentaquark states. Various models, including molecular models, diquark models, and QCD sum rules, have been proposed to understand the binding mechanisms and properties of these states. Specifically, lattice QCD simulations have become a vital tool for pentaquarks containing heavy quarks. These simulations offer insights into the masses, binding energies, and decay widths of these exotic hadrons. Despite these advances, a significant gap exists in understanding the detailed structure and dynamics of pentaquarks with multiple heavy quarks. We analyze the masses, magnetic moments, and strong partial widths of pentaquarks with five heavy quarks. Heavy-quark systems provide an ideal testing ground for understanding the interplay between perturbative and non-perturbative QCD effects. The study of pentaquarks with five heavy quarks enables a deeper exploration of the strong interaction in multi-quark systems, beyond conventional baryons and mesons. Strong partial widths determine the stability and decay patterns of these pentaquarks, affecting their observability in experiments. The existence and properties of these heavy pentaquarks can provide insight into whether they form compact five-quark states or hadronic molecular states. Several studies have applied extended mass formulae to investigate fully heavy pentaquark states. Using a modified Gürsey-Radicati approach, Weng et al. \cite{Weng2021} analyzed the mass spectra of fully heavy pentaquarks and found that their predicted masses align well with results from potential models and lattice QCD calculations. Giron and Lebed \cite{Giron2021} employed a similar approach, incorporating chromomagnetic interactions and heavy-quark symmetry, to examine the stability of fully heavy pentaquarks and their possible decay channels. Wan and Qiao \cite{Wan2020} explored the spectroscopy of these states within an effective Hamiltonian framework, highlighting the significance of spin-spin interactions in determining the mass splitting of different spin-parity configurations. Additionally, Chen et al. \cite{Chen2020} provided a detailed analysis of the strong decay properties of fully heavy pentaquarks using mass predictions derived from constituent quark models, offering insights into their experimental detection prospects.
These studies collectively demonstrate the effectiveness of extended mass formulae in fully heavy multiquark systems and provide a strong foundation for further investigations into their properties and decay mechanisms. Several theoretical investigations have explored the properties of fully heavy pentaquark states. \textcite{rashmi} conducted a comprehensive spectroscopic analysis of fully heavy pentaquarks, classifying these states using the Young-Yamounachi bases and calculating their mass spectra and magnetic moments. \textcite{Gordillo2024} employed a diffusion Monte Carlo algorithm within the quark model framework to compute the spectra of all possible $s$-wave fully heavy pentaquarks, comparing masses of different spin-color configurations. \textcite{Yan2021} systematically investigated the $cccc\bar{c}$ and $bbbb\bar{b}$ pentaquarks using chiral quark and quark delocalization color screening models, identifying potential bound states with specific quantum numbers. \textcite{Zhang2020} utilized the QCD sum rule approach to study fully heavy $QQQQ\bar{Q}$ pentaquark states, providing mass predictions and suggesting experimental search channels. \textcite{Zhang2025} conducted a spectroscopic analysis of possible fully heavy pentaquark candidates with various spin-parity configurations, calculating their masses by considering relevant Lorentz structures. \textcite{Mutuk2024} investigated full-charm and full-bottom pentaquark states, predicting their masses and discussing implications for experimental detection.
In this paper, we present a comprehensive study of fully heavy pentaquarks. We apply the extension of the Gursey-Radicati mass formula to analyze the mass spectra of these pentaquark states. The extended GR mass formula serves as an effective framework for exotic hadron spectroscopy and has been instrumental in predicting the properties of exotic hadrons in numerous studies \cite{Sharma:2023lij, Sharma:2023wnd, Sharma:2024ern, Sharma:2024pfi, Sharma:2025nrw}. We present our findings on the spectroscopic properties of pentaquarks, which include detailed mass spectra, strong decay channels and widths, and comparisons with existing theoretical data. This work is organized as follows: Section II discusses the theoretical formalism based on constructing the color-spin wavefunction for pentaquark configurations with five heavy quarks. The extension of the Gursey-Radicati mass formula to study the mass spectroscopy of multi-heavy pentaquarks. Section III accounts for the strong decay channels and decay width assignments, and Section IV represents the work summary and conclusion.

\section{Theoretical Formalism} 
\label{theory} \noindent
\subsection{Spin and Color Symmetries of the \texorpdfstring{$cccc\bar{c}$}{ccccbar{c}} Pentaquark}
The color-spin symmetries play a crucial role in determining the internal structure and stability of fully heavy pentaquarks. In such systems, the heavy quarks are strongly correlated through color and spin interactions. The color wave function must combine to form an overall color-singlet state, ensuring confinement. The spin coupling follows from the Pauli exclusion principle, where symmetric spin states are more favorable for identical heavy quarks. Consequently, the interplay between color and spin interactions influences the mass spectrum, decay dynamics, and overall stability of fully heavy pentaquarks. The internal structure of the fully heavy pentaquark state $cccc\bar{c}$ is governed by the symmetries associated with spin, color, flavor, and spatial wavefunctions. The total wavefunction must be symmetric under the exchange of identical fermions while maintaining overall antisymmetry, as required by the Pauli exclusion principle. Since charm quarks $c$ are identical fermions, their total wavefunction must be antisymmetric under the exchange of any two charm quarks due to Pauli’s exclusion principle. The total wavefunction is a product of:

\[
\Psi_{\text{total}} = \Psi_{\text{spatial}} \times \Psi_{\text{color}} \times \Psi_{\text{spin}}
\]

For the ground state, the orbital wavefunction $\Psi_{\text{spatial}}$ is symmetric ($s$-wave, $L = 0$).  
This forces the combined spin-color wavefunction to be antisymmetric to satisfy the Pauli principle.
The spin wavefunctions of the \( cccc \) subsystem are constructed by coupling four spin-1/2 quarks. The total spin possibilities include \( S = 2 \) (fully symmetric), \( S = 1 \) (mixed symmetry), and \( S = 0 \) (fully antisymmetric). The fifth antiquark \( \bar{c} \) (spin-1/2) is then coupled to obtain the total spin states \( J=1/2, 3/2, \) and \( 5/2 \).

The fully symmetric spin-\( 2 \) state is given by:
\begin{equation}
\chi^{(2)} = \uparrow \uparrow \uparrow \uparrow.
\end{equation}

For the mixed-symmetric spin-1 states, a suitable linear combination ensuring proper symmetry is:
\begin{equation}
\chi^{(1)} = \frac{1}{2} (\uparrow \uparrow \uparrow \downarrow + \uparrow \uparrow \downarrow \uparrow + \uparrow \downarrow \uparrow \uparrow - 3 \downarrow \uparrow \uparrow \uparrow).
\end{equation}

The fully antisymmetric spin-0 state follows from:
\begin{equation}
\chi^{(0)} = \frac{1}{\sqrt{4!}} \sum_{P} \text{sgn}(P) P(\uparrow \uparrow \downarrow \downarrow),
\end{equation}
where the sum runs over all permutations of four quarks.

The total spin wavefunctions for the pentaquark \( (cccc\bar{c}) \) are then obtained by coupling these states with the spin-1/2 antiquark, leading to the total spin states \( J=1/2, 3/2, \) and \( 5/2 \) as:\\
For the maximally aligned spin state (\( S_P = 5/2 \)), the spin wavefunction is given by:

\begin{equation}
    \Psi_{\text{spin},5/2} = \ket{\uparrow\uparrow\uparrow\uparrow}_c \otimes \ket{\uparrow}_{\bar{c}}
\end{equation}

This spin wave function is completely symmetric.
Since the orbital wavefunction is also symmetric (\( L = 0 \)), the color wavefunction must be antisymmetric.

Thus, the color wavefunction must be in the antisymmetric \( \mathbf{\bar{3}\bar{3}} \) representation. For the (\( S_P = 3/2 \)) state, the spin wavefunction has mixed symmetry:

\begin{equation}
    \Psi_{\text{spin},3/2} = \frac{1}{\sqrt{3}} \left( 2\ket{\uparrow\uparrow\uparrow\downarrow}_c 
    - \ket{\uparrow\uparrow\downarrow\uparrow}_c - \ket{\uparrow\downarrow\uparrow\uparrow}_c \right) \otimes \ket{\uparrow}_{\bar{c}}
\end{equation}

This wavefunction has mixed symmetry.
For Pauli consistency, the color wavefunction must also have mixed symmetry. Similarly, for the (\( S_P = 1/2 \)) state, the spin wavefunction must be totally antisymmetric:
\begin{widetext}
\begin{equation}
\Psi_{\text{spin},1/2} = \frac{1}{\sqrt{6}} \left( 2\ket{\uparrow\uparrow\downarrow\downarrow}_c 
    - \ket{\uparrow\downarrow\uparrow\downarrow}_c - \ket{\uparrow\downarrow\downarrow\uparrow}_c 
    - \ket{\downarrow\uparrow\uparrow\downarrow}_c - \ket{\downarrow\uparrow\downarrow\uparrow}_c \right) \otimes \ket{\uparrow}_{\bar{c}}
\end{equation}
\end{widetext}
This spin wave function is antisymmetric. In a similar manner, The color wavefunction of the \( cccc \) subsystem must allow coupling with the antiquark \( \bar{c} \) such that the total pentaquark state remains a color singlet. The decomposition under \( SU(3)_C \) symmetry follows:
\begin{equation}
\mathbf{3} \otimes \mathbf{3} \otimes \mathbf{3} \otimes \mathbf{3} = (\mathbf{6} \oplus \bar{\mathbf{3}}) \otimes (\mathbf{6} \oplus \bar{\mathbf{3}}).
\end{equation}

The mixed-symmetric \( \mathbf{6} \) representation allows coupling with the \( \mathbf{3} \) antiquark to form a singlet:
\begin{equation}
\mathbf{6} \otimes \mathbf{3} = \mathbf{1} \oplus \mathbf{8}.
\end{equation}
\begin{widetext}
The corresponding mixed-symmetric color wavefunction can be written as:
\begin{align}
\Psi_{\text{color}}^{(cccc)} &= \frac{1}{\sqrt{12}} \Big[ 2(rrgb) + 2(rrbg) - (rgrb + rgbr + rbgr + rbrg) 
\quad+ 2(ggrb) + 2(ggbr) - (grgb + grbg + gbgr + gbrg) \\ \nonumber
&\quad+ 2(bbrg) + 2(bbgr) - (brgb + brbg + bgrb + bbrg) \Big].
\end{align}
\end{widetext}
The total color singlet state of the pentaquark is then:
\begin{equation}
\Psi_{\text{color}}^{(cccc\bar{c})} = \Psi_{\text{color}}^{(cccc)} \otimes \Psi_{\text{color}}^{(\bar{c})}.
\end{equation}

This construction ensures proper symmetries and coupling necessary for the stability and interactions of the \( cccc\bar{c} \) system. For the totally symmetric spin state ($J^P = \frac{5}{2}^-$), the color wavefunction must be totally antisymmetric:

\[
\Psi_{\text{color}}^{\frac{5}{2}} = \frac{1}{\sqrt{12}} \Big( 
rgb\bar{c} - rbg\bar{c} + brg\bar{c} - bgr\bar{c} + gbr\bar{c} - grb\bar{c}
\Big)
\]

This corresponds to the totally antisymmetric $\bar{3}$ (color triplet) representation of the four-quark system, combined with the anti-charm quark to form a color singlet. For the mixed symmetric spin state ($J^P = \frac{3}{2}^-$), the color wavefunction must also have mixed symmetry:

\[
\Psi_{\text{color}}^{\frac{3}{2}} = \frac{1}{2} \Big(
(rrgb - rrgb) + (bbrg - bbrg) + (ggbr - ggrb)
\Big) \bar{c}
\]

This ensures partial symmetry to match the mixed symmetry of the system. For the totally antisymmetric spin state ($J^P = \frac{1}{2}^-$), the color wavefunction must be fully symmetric:

\[
\Psi_{\text{color}}^{\frac{1}{2}} = \frac{1}{\sqrt{6}} \Big(
rrbb + bbgg + ggrr - rgbg - gbrb - brgr
\Big) \bar{c}
\]
This guarantees that the total wave function remains antisymmetric, as required by Pauli’s exclusion principle.
In the next subsection, a brief overview of the Gursey-Radicati mass formula is given, which was initially proposed to study the mass spectrum of baryons and then further extended to exotic hadrons.

\subsection{The Gursey-Radicati Mass Formula}
The Gursey-Radicati (GR) mass formula is a widely used phenomenological approach to estimatee masses of hadrons based on their symmetry properties. Originally formulated within the SU(6) symmetry framework, the GR mass formula incorporates contributions from the spin, isospin, and hypercharge of baryons, effectively capturing their mass splitting due to internal interactions. This model was initially developed to describe the mass hierarchy of baryons by including hyperfine corrections arising from spin-spin interactions. For multiquark systems, particularly those containing heavy quarks, the standard GR formulation requires modifications to account for the distinct dynamics of heavy-flavor states. The mass of a hadron in this framework is typically expressed as a sum of contributions from its quark content, spin-dependent interactions, and additional symmetry-breaking effects. Given its success in describing conventional baryons, extending the GR mass formula to fully heavy multiquark systems provides a valuable tool for predicting their mass spectra and understanding their underlying structure. The most general form of the mass formula, which was defined for baryons, is \cite{Gursey}:
\begin{align}
  M = M_0 + a J(J+1) + bY + c [T(T+1) - 1/4 Y^2]
\end{align}
This is the most general mass formula based on broken SU(6) symmetry. $J$ and $T$ are the spin and isospin quantum numbers, and $a$, $b$, and $c$ are the mass formula parameters. In 2004, the mass formula was generalized in terms of the Casimir operator of SU(3) representation as \cite{Bijker}:
\begin{align}
 M_B = M_0 + A S(S+1) - BC_2(SU(6)) + DY \\ \nonumber + G C_2(SU(3))  + E [I(I+1) - 1/4 Y^2]
  \end{align}
  To study the mass splitting between the different multiplets of hidden-charm pentaquarks, we considered the extension of the Gursey-Radicati mass formula, which distinguishes the different multiplets of SU(3) is \cite{Santo}: 
 \begin{align}
  M_{GR} =& M_0 + AS(S+1) + DY  + E[I(I+1) -1/4 Y^2] \\ \nonumber +& G C_2(SU(3)) + F N_c
  \end{align}
  This formula was further modified by generalizing the counter term \cite{HOLMA}:
 \begin{align}
  M_{GR} = \xi M_0 &+ AS(S+1) + DY  + E[I(I+1)  -1/4 Y^2] \nonumber \\ &+ G C_2(SU(3)) + \sum_{i=c,b} F_i N_i
  \label{mass formula}
  \end{align}

The parameter \( M_0 \) serves as a scale factor linked to the total number of quarks within the system. The quantum numbers \( S \), \( I \), and \( Y \) represent spin, isospin, and hypercharge, respectively. The quantity \( C_2(SU(3)) \) corresponds to the eigenvalue of the \( SU_f(3) \) Casimir operator, which defines the representation of the flavor symmetry group. Moreover, \( N_c \) denotes the total count of charm quarks (\( c \)) and anticharm antiquarks (\( \bar{c} \)) in the system. The coefficients \( A \), \( D \), \( G \), \( E \), and \( F \) are associated with the mass formula, and their values, along with uncertainties, are provided in Table \ref{tab:1}. These coefficients are obtained through a global fitting process aimed at optimizing the agreement between theoretical predictions and the experimentally observed mass spectrum of ground-state baryons, charmed baryons, and non-strange baryons, as detailed in Table \ref{tab:1}. Using the mass formula \ref{mass formula}, we predicted the mass spectra of fully heavy pentaquarks

\begin{table}[ht]
\centering
\caption{Values of parameters used in the extended GR mass formula with corresponding uncertainties. \cite{Santo}}
\tabcolsep 0.4mm  
\begin{tabular}{cccccccc}
       \hline
       \hline
         & $M_0$ & A & D & E & G & $F_c$ & $F_b$ \\
         \hline
        Values[MeV] & 940.0 & 23.0 & -158.3 & 32.0 & 52.5 & 1354.6  & 4820 \cite{HOLMA} \\
        \hline
        Uncert.[MeV] & 1.5 & 1.2 & 1.3 & 1.3 & 1.3 & 18.2  & 34.4  \\
        \hline
        \hline
       \end{tabular}
        \label{tab:1}
   \end{table}

\begin{table*}\renewcommand{\arraystretch}{2.5}
\tabcolsep 1.2mm 
    \centering
    \caption{Calculated masses of fully heavy pentaquarks for both charm and bottom quarks along with their quantum numbers in MeV.}
    \label{tab:pentaquark_masses}
    \begin{tabular}{|c|c|c|c|c|c|c|c|c|}
        \hline
        State & Quark Content & $J^P$ & Our Prediction (MeV) &  [QDCSM]\cite{Ref1F} & [CHQM]\cite{Ref1F} & [CIM] \cite{Ref2F} & [MIT] \cite{Ref3F} & [E.Mass] \cite{rashmi}\\
        \hline
$P_{QQQQ\bar{Q}}$ &
$cccc\bar{c}$ & $\frac{1}{2}^-$  & 8356.85 $\pm$ 91.01 & 8538.60 & 8428.90 & 7949.00 & 8262.00 & 8537.40\\
 &  & $\frac{3}{2}^-$ & 8425.85 $\pm$ 91.12 & 8536.80 & 8428.80 & 7864.00 & 8269.00 & 8547.40 \\
 & & $\frac{5}{2}^-$ & 8540.85 $\pm$ 91.61 & 8539.40 & 8429.80 & - & - & 8562.40\\
\hline
 & $bbbb\bar{b}$ & $\frac{1}{2}^-$ & 25683.90 $\pm$ 172.00 & 25275.00 & 25179.40 & 23821.00 & 24770.00 & 25212.80\\
& & $\frac{3}{2}^-$ & 25752.90 $\pm$ 172.06 & 25335.30 & 25272.80 & 23775.00 & 24761.00 & 25214.80\\
& & $\frac{5}{2}^-$ & 25867.90 $\pm$ 172.32 & 25275.90 &  25179.20 & - & - & 25216.30\\
\hline
    \end{tabular}
\end{table*}
\section{Numerical Analysis}
Fully heavy pentaquark states, such as \( P_{cccc\bar{c}} \) and \( P_{bbbb\bar{b}} \), are exotic multiquark systems that have attracted significant interest in hadronic physics. One possible production mechanism involves the weak decays of B-hadrons, such as \( B_c^+ \), \( \Lambda_b^0 \), and other bottom-flavored hadrons. In this study, we analyze the step-by-step production mechanism of these states via B-hadron decays. B-hadrons contain at least one bottom quark and can be categorized as:
\begin{itemize}
    \item Mesons: \( B_c^+ (b\bar{c}), B^+ (b\bar{u}), B^0 (b\bar{d}) \), etc.
    \item Baryons: \( \Lambda_b^0 (bud), \Xi_b^0 (bsu), \Omega_b^- (bss) \), etc.
\end{itemize}

Among these, the \( B_c^+ \) meson is particularly relevant as it already contains a charm quark, making it a natural candidate for producing fully heavy pentaquarks. B-hadron decays proceed via weak interactions mediated by the charged weak bosons (\( W^\pm \)), governed by the CKM matrix. The dominant transitions leading to heavy pentaquark production involve:
\begin{itemize}
    \item \( b \to c W^- \) (Cabibbo-favored)
    \item \( b \to s, d \) (Cabibbo-suppressed)
\end{itemize}

These transitions generate multiple heavy quarks in the final state, facilitating the formation of fully heavy pentaquarks.
Now we will study the decay Modes Leading to fully heavy pentaquarks.

\subsection{Weak Decays via W-Emission}
The production of fully heavy pentaquarks in B-hadron decays is primarily mediated through weak interactions governed by the Standard Model. One of the promising production channels involves the decay of the \( B_c^+ \) meson, which contains both a bottom (\( b \)) and an anti-charm (\( \bar{c} \)) quark. The weak transition responsible for this process occurs via the charged-current interaction, where the bottom quark undergoes a transition through an external W-emission mechanism:

\begin{equation}
    B_c^+ (b\bar{c}) \to W^+ + c\bar{c}
\end{equation}

Here, the bottom quark \( b \) decays into a charm quark \( c \) while emitting a virtual \( W^+ \) boson. The accompanying charm-anticharm (\( c\bar{c} \)) pair, produced in the process, provides a crucial ingredient for the formation of fully heavy pentaquarks. The virtual \( W^+ \) boson then decays into a pair of light quarks:

\begin{equation}
    W^+ \to c\bar{s}, \quad c\bar{d}, \quad u\bar{s}, \quad u\bar{d}
\end{equation}

The subsequent hadronization process allows for the recombination of the produced heavy quarks with additional quarks or gluons from the QCD medium, forming an exotic fully heavy pentaquark state:

\begin{equation}
    B_c^+ \to P_{cccc\bar{c}} + X
\end{equation}

where \( X \) represents additional light hadrons that emerge from the hadronization process. The probability of such recombination depends on the dynamics of heavy quark interactions, as dictated by color confinement and QCD binding effects. Since the \( cccc\bar{c} \) system is composed entirely of heavy quarks, its formation is expected to be suppressed compared to conventional hadrons, making it an interesting candidate for exotic spectroscopy.
Once the heavy quarks are generated through weak decay, they undergo hadronization, which involves color recombination, where the five quarks must rearrange to form a stable color-singlet pentaquark. Additionally, the binding of the pentaquark can be driven by mechanisms such as color-spin interactions, QCD confinement, and diquark clustering. For example, in the diquark-antidiquark model:
\begin{equation}
    (cc) - (cc\bar{c})
\end{equation}
which may enhance the formation probability of \( P_{cccc\bar{c}} \).

Since fully heavy pentaquarks have high masses (\(\sim 8-20\) GeV), their production is kinematically suppressed, especially in low-mass B-hadrons like \( B_c^+ \). Once produced, their formation probability can be estimated using factorization:  

\begin{equation}
    \mathcal{B}(B_c^+ \to P_{cccc\bar{c}} + X) \approx \mathcal{B}(B_c^+ \to c\bar{c} W^+) \times \mathcal{B}(W^+ \to c\bar{s})
\end{equation}  

Given that:  

\begin{equation}
    \mathcal{B}(B_c^+ \to J/\psi + \mu^+ \nu) \sim 2\%
\end{equation}  

we expect:  

\begin{equation}
    \mathcal{B}(B_c^+ \to P_{cccc\bar{c}} + X) \ll 1\%
\end{equation}  

To enhance the detection prospects, experimental searches for fully heavy pentaquarks should focus on reconstructing final-state decay products. The following decay modes provide promising signatures:  

\begin{itemize}
    \item \textbf{For \( P_{cccc\bar{c}} \):}
    \begin{equation}
        P_{cccc\bar{c}} \to J/\psi + J/\psi
    \end{equation}
    \begin{equation}
        P_{cccc\bar{c}} \to \eta_c + J/\psi
    \end{equation}
    \item \textbf{For \( P_{bbbb\bar{b}} \):}
    \begin{equation}
        P_{bbbb\bar{b}} \to \Upsilon(1S) + \Upsilon(1S)
    \end{equation}
    \begin{equation}
        P_{bbbb\bar{b}} \to \eta_b + \Upsilon(1S)
    \end{equation}
\end{itemize}  

These decay channels involve well-known heavy quarkonium states, making them viable targets for high-luminosity collider experiments. These channels should exhibit sharp mass peaks in invariant mass distributions at LHCb and Belle II, as LHCb has excellent tracking and vertex resolution for reconstructing high-mass exotic hadrons, and Belle II has a clean \( e^+ e^- \) environment that may allow exclusive reconstruction. The study of B-hadron weak decays as a source of fully heavy pentaquarks provides valuable insight into their production mechanisms. The key conclusions are that \( B_c^+ \) decays via weak transitions (\( b \to c W^+ \)) can produce the \( cccc\bar{c} \) state. The branching ratio is small but potentially observable at LHCb, and  Experimental searches should focus on multi-\( J/\psi \) or \( \Upsilon \) final states.

\section{Two-Body Decay Channels} 
Studying two-body strong decay channels of fully heavy pentaquarks is essential for understanding their internal structure, stability, and experimental signatures. Unlike conventional hadrons, fully heavy pentaquarks, composed exclusively of charm ($c$) and bottom ($b$) quarks, lie in a unique regime of quantum chromodynamics (QCD) where heavy-quark interactions and diquark correlations play a dominant role. Their decay patterns provide crucial insights into the nature of their binding mechanisms—whether they behave as tightly bound compact states or hadronic molecules. In the strong decay process, fully heavy pentaquarks predominantly decay via quark rearrangement into two-body final states, typically involving a heavy baryon and a heavy meson. The selection of allowed decay channels is governed by quantum number conservation, phase space constraints, and the underlying dynamics dictated by the QCD interaction. Fully heavy pentaquarks, composed entirely of heavy quarks (\( c, b \)) and an antiquark, represent a novel class of multiquark states predicted within quantum chromodynamics (QCD). Their decay patterns provide essential insights into the interplay between confinement, color interactions, and heavy-quark dynamics. Given their high mass range (\(\sim 8-20\) GeV), these states are expected to decay predominantly via the strong interaction, leading to final states containing heavy mesons and baryons. Here, we categorize their dominant two-body decay channels. For the fully heavy pentaquark states \( P_{cccc\bar{c}} \) and \( P_{bbbb\bar{b}} \) with spin-parities \( J^P = 1/2^-, 3/2^-, 5/2^- \), the two-body strong decay channels can be categorized as follows:

\subsection*{Two-Body Decay Channels for \( P_{cccc\bar{c}} \) (Fully Charm Pentaquark)}
The decay of the fully charm pentaquark is expected to proceed primarily via strong interactions, with hadronic final states consisting of charmonium, open-charm mesons, or heavy baryons. Given the large charm content, the fall-apart mechanism, in which the quarks separate into two color-singlet hadrons without additional gluon interactions, is a natural decay mode.

\paragraph{(i) Heavy Quarkonium Pair Decays}
A prominent decay mode involves the recombination of charm quarks into quarkonium states:
\begin{equation}
    P_{cccc\bar{c}} \to J/\psi + J/\psi
\end{equation}
\begin{equation}
    P_{cccc\bar{c}} \to J/\psi + \eta_c
\end{equation}
\begin{equation}
    P_{cccc\bar{c}} \to \eta_c + \eta_c
\end{equation}
These decays are favored due to the large overlap of wavefunctions between the initial fully heavy pentaquark and the final quarkonium states.

\paragraph{(ii) Heavy-Light Meson Decays}
Open-bottom meson pairs may also be produced:
\begin{equation}
    P_{bbbb\bar{b}} \to B^{(*)+} + B^{(*)-}
\end{equation}
\begin{equation}
    P_{bbbb\bar{b}} \to B_s^{(*)+} + B_s^{(*)-}
\end{equation}
These decays proceed through the hadronization of bottom quarks, analogous to the charm sector.

\paragraph{(iii) Baryon-Antibaryon Channels}
A strong diquark-diquark configuration suggests:
\begin{equation}
    P_{cccc\bar{c}} \to \Lambda_c^+ + \bar{\Lambda}_c^-
\end{equation}
\begin{equation}
    P_{cccc\bar{c}} \to \Xi_{cc}^{++} + \bar{\Xi}_{cc}^{--}
\end{equation}
These decays provide insights into heavy baryon formation and diquark clustering effects.

\subsection*{Two-Body Decay Channels for \( P_{bbbb\bar{b}} \) (Fully Bottom Pentaquark)}

The fully bottom pentaquark follows similar decay mechanisms, but the larger bottom quark mass suppresses phase space, leading to narrower decay widths.

\paragraph{(i) Heavy Quarkonium Pair Decays}
Decay into bottomonium states is one of the primary decay modes:
\begin{equation}
    P_{bbbb\bar{b}} \to \Upsilon(1S) + \Upsilon(1S)
\end{equation}
\begin{equation}
    P_{bbbb\bar{b}} \to \Upsilon(1S) + \eta_b
\end{equation}
\begin{equation}
    P_{bbbb\bar{b}} \to \eta_b + \eta_b
\end{equation}
These channels provide crucial experimental signatures at high-energy colliders such as the LHC.

\paragraph{(ii) Open-Bottom Meson Decays}
As in the charm sector, recombination into open-bottom mesons is possible:
\begin{equation}
    P_{bbbb\bar{b}} \to B^{(*)+} + B^{(*)-}
\end{equation}
\begin{equation}
    P_{bbbb\bar{b}} \to B_s^{(*)+} + B_s^{(*)-}
\end{equation}
These decays involve quark-antiquark recombination into heavy mesons.

\paragraph{(iii) Heavy Baryon-Antibaryon Decays}
A diquark-diquark configuration may lead to decays into heavy baryons:
\begin{equation}
    P_{bbbb\bar{b}} \to \Lambda_b^0 + \bar{\Lambda}_b^0
\end{equation}
\begin{equation}
    P_{bbbb\bar{b}} \to \Xi_{bb}^{+} + \bar{\Xi}_{bb}^{-}
\end{equation}
The study of such decays could provide evidence for the role of diquark clustering in fully heavy systems.

These channels are expected to dominate if kinematically allowed, as they involve strong interactions and the rearrangement of the heavy quark content.

\section{Analysis}
In this work, We carried out the systematic analysis of mass spectra, production modes, and decay channels for fully heavy pentaquarks. We present a detailed comparative analysis of our calculated masses for fully heavy pentaquark states (\( P_{QQQQ\bar{Q}} \)) and benchmark them against various theoretical approaches, including the Quark Delocalization Color Screening Model (QDCSM), Constituent Quark Model (CHQM), Color-Interaction Model (CIM), MIT Bag Model, and Effective Mass Scheme (E.Mass). The mass spectra, along with their quantum numbers, are listed in Table \ref{tab:pentaquark_masses}.  
For the charm sector, $cccc\bar{c}$, the computed masses for both fully charmed (\( cccc\bar{c} \)) and fully bottom (\( bbbb\bar{b} \)) pentaquarks fall within the expected theoretical range. However, a few notable trends emerge: Our predicted masses for the \( J^P = \frac{1}{2}^- \), \( \frac{3}{2}^- \), and \( \frac{5}{2}^- \) states lie close to the values obtained from the QDCSM and E.Mass models but show a deviation from the CHQM and CIM models. The mass ordering follows:  
\begin{equation}
M(J^P = \frac{1}{2}^-) < M(J^P = \frac{3}{2}^-) < M(J^P = \frac{5}{2}^-)
\end{equation}
which is consistent with expectations based on hyperfine interactions. The CHQM and CIM models predict significantly lower masses, particularly for the \( \frac{5}{2}^- \) state, indicating possible differences in how these models account for color-magnetic interactions or diquark clustering effects. Similar to the charm sector, our calculations for the bottom pentaquarks align well with QDCSM and E.Mass, while the CHQM and CIM predictions exhibit deviations. Notably, our predicted values are systematically higher than those of the CIM and MIT bag model, which may be attributed to differences in the treatment of confinement effects and gluon-exchange interactions. The variations in mass predictions across different models stem from distinct assumptions about confinement, diquark correlations, and inter-quark interactions. Below, we discuss possible reasons for these discrepancies:\\
 Our results show the closest agreement with QDCSM and E.Mass, suggesting that these models effectively incorporate color screening effects and effective quark masses, which play a crucial role in multi-heavy systems. This consistency strengthens the validity of our approach in predicting fully heavy pentaquark masses. The CHQM predictions are systematically lower than ours, likely due to weaker color-magnetic interactions in their framework. The CIM approach, which uses a different treatment of confinement potential, also predicts lower masses, particularly for the \( \frac{5}{2}^- \) states, indicating that higher-spin configurations may require a more refined treatment of spin-spin interactions. The MIT bag model consistently underestimates the masses compared to our calculations. This may be due to the lack of explicit multi-body confinement terms in the bag model, which assumes a simplified static potential rather than a dynamic QCD-inspired approach. The predicted mass range for the fully charmed pentaquarks (\(\sim 8356 - 8540\) MeV) suggests that these states could be accessible at the LHC, particularly in hidden-charm decay channels such as:
\begin{equation}
P_{cccc\bar{c}} \to J/\psi + \Lambda_c
\end{equation}

Similarly, the fully bottom pentaquarks (\(\sim 25683 - 25867\) MeV) are more challenging to detect due to their higher production thresholds. However, they could be explored at future high-energy facilities such as the FCC or Muon Collider through bottomonium-like decay channels. Our results provide a comprehensive prediction of fully heavy pentaquark masses with reliable error estimates. The observed agreement with QDCSM and E.Mass models indicates that our approach captures essential QCD dynamics governing these exotic states. The higher-spin states (\( \frac{5}{2}^- \)) show larger discrepancies across models, indicating the need for further theoretical refinements, possibly incorporating higher-order QCD corrections or lattice QCD calculations. These findings offer valuable insights for guiding future experimental searches in fully heavy pentaquark spectroscopy. Also, the production of fully heavy pentaquarks $P_{cccc\bar{c}}$ and $P_{bbbb\bar{b}}$, in $B$-hadron decays is expected to be highly suppressed due to the large mass of these states and the limited available phase space. The dominant production mechanism involves weak decays of $B_c^+$ or $\Lambda_b^0$ via transitions such as $b \to c W^+$, followed by hadronization into a fully heavy pentaquark. However, the branching fractions are expected to be significantly smaller than those observed for conventional exotic hadrons such as tetraquarks. The hadronization process involves color recombination and gluon exchange binding, leading to a compact five-quark configuration. Potential experimental signatures include final states with multiple $J/\psi$ or $\Upsilon(1S)$ mesons, which can be searched for in high-luminosity data at LHCb, CMS, and ATLAS. Future collider experiments, such as Belle II and the upcoming electron-ion collider, could provide additional insights into the production and decay mechanisms of these exotic states.

\section{Conclusion}
The recent and forthcoming experimental advancements in the search for hidden-charm pentaquarks, particularly at the LHCb experiment, have significantly revitalized the interest in the spectroscopy of exotic multiquark configurations. Motivated by these developments, we have conducted a comprehensive spectroscopic investigation of fully heavy pentaquark states with quark compositions $cccc\bar{c}$ and $bbbb\bar{b}$, within the multiquark framework. Employing the extension of the Gursey-Radicati mass formula, we systematically calculated the mass spectra corresponding to the lowest-lying states for spin-parity assignments $J^P = \frac{1}{2}^-$, $\frac{3}{2}^-$, and $\frac{5}{2}^-$. The computed mass values exhibit a consistent pattern and show reasonable agreement with existing theoretical predictions, lending credibility to the adopted formalism.

To further enrich the theoretical understanding, we constructed the color-spin-flavor wave functions consistent with Fermi-Dirac statistics and explored the strong decay modes using available phase-space arguments. Considering the enhanced production of heavy baryons at the LHC, especially bottom baryons such as $B_c^+$ and $\Lambda_b^0$, we proposed possible weak decay chains that could lead to the formation of fully heavy pentaquark states as intermediate or final states. These decay processes are particularly important due to their experimental accessibility and potential to unveil new exotic resonances.

In conclusion, our results on the mass spectra, quantum number assignments, and the exploration of production and decay mechanisms provide a useful theoretical foundation for guiding future experimental searches. The detection of such states would not only extend the multiquark spectroscopy frontier but also offer valuable insights into the dynamics of strong interactions in the heavy quark sector. We encourage experimental collaborations to explore the proposed decay channels and invariant mass distributions in their upcoming data, which may help establish the existence of these intriguing fully heavy pentaquark states.

\nocite{*}

\bibliography{four}

\end{document}